\documentclass{llncs}
\usepackage{xspace}
\usepackage{wrapfig}
\usepackage{paralist}
\usepackage{stex-logo}
\usepackage{tikz}
\usetikzlibrary{positioning,arrows,shapes,fit,snakes}
\usepackage[today,eso-foot]{svninfo}
\svnInfo $Id: paper.tex 1250 2012-04-22 09:44:58Z kohlhase $
\svnKeyword $HeadURL: https://svn.mathweb.org/repos/planetary/doc/cicm12/paper.tex $
\usepackage[style=alphabetic,isbn=false]{biblatex}
\renewbibmacro*{event+venue+date}{}

\bibliography{kwarc}
\def\pn{\textsc{Planetary}\xspace}
\def\pnp{\pn project\xspace}
\def\pns{\pn system\xspace}
\def\infobar{\texttt{InfoBar}\xspace}
\def\foldingbar{\texttt{FoldingBar}\xspace}

\title{The \protect\pn Project: Towards eMath3.0}
\author{Michael Kohlhase -- for the \protect\pn Group}
\authorrunning{Kohlhase}
\institute{
\begin{tabular}[t]{c}
  Computer Science, Jacobs University Bremen, Germany
\end{tabular}\\ 
\protect\url{http://planetary.mathweb.org}
}

\begin{document}
\maketitle
\begin{abstract}
  The \pnp develops a general framework -- the {\pns} -- for social semantic portals that
  support users in interacting with STEM (Science/Technology/Engineering/Mathematics)
  documents. Developed from an initial attempt to replace the aging portal of
  PlanetMath.org with a mashup of existing MKM technologies, the \pns is now in a state,
  where it can serve as a basis for various eMath3.0 portals, ranging from eLearning
  systems over scientific archives to semantic help systems.
\end{abstract}

The \pnp aims at developing a general framework -- the {\pns} -- for social semantic
portals that support users in interacting with STEM documents. It is carried by
enthusiasts from Jacobs University and The Open University.

\subsection*{Main Concepts and Project Genesis}
Work on the \pns was triggered in August 2010 by the realization that the KWARC group at
Jacobs University had developed semantic counterparts of much of the components underlying
the PlanetMath portal~\cite{planetmath:on}. PlanetMath.org is an online community that
creates and manages an encyclopedia of mathematical concepts; hundreds of regular
contributors have published about 8500 encyclopedia entries called articles. PlanetMath
was founded in 2000; even before Wikipedia, and is thus one of the first Web2.0 systems.
The No\"osphere system~\cite{noosphere:on} underlying the portal -- essentially a
\LaTeX-based Wiki implemented in Perl -- is showing its age and becoming hard to manage.
We felt that extending Planetmath to an eMath3.0 system -- a social semantic web platform
for Mathematics -- via MKM technologies, might breathe additional life into the PlanetMath
community and at the same time serve as a showcase of MKM technologies into the
mathematics community.

The pre-existing MKM components that can be combined to form a semantic counterpart of
No\"osphere are (see also Figure~\ref{fig:arch}):
\begin{compactenum}
\item TNTBase~\cite{tntbase:trac} for web-enabled, versioned
  storage
\item The LaTeXML daemon~\cite{GinStaKoh:latexmldaemon11} for
  transforming {\TeX/\LaTeX} document fragments to HTML5\footnote{We use HTML5 as it
    integrates HTML for document layout with MathML for formula presentation, SVG for
    diagrams, and RDFa for document-embedded metadata and is supported by the major
    browsers.}
\item \sTeX~\cite{sTeX:online}, a semantic variant of {\LaTeX} that can
  be transformed to OMDoc~\cite{Kohlhase:OMDoc1.2} and further to semantically annotated
  HTML5~\cite{KMR:NoLMD08}
\item JOBAD~\cite{JOBAD:on}, a JavaScript API embedding semantic
  services into Web documents
\end{compactenum}
The only missing piece was a front-end that integrated them, added user and permissions
management, and added discussion fora (an essential feature of PlanetMath). We found this
component in the open source Vanilla Forums system~\cite{VanillaForums}, that could easily
be extended by wrapping the MKM components into Vanilla plugins. Another plugin that had
to be added to the mix for PlanetMath feature parity was a system for metadata management
and visualization: \pn exports metadata to an RDF triple store (here an instance of the
Openlink Virtuoso system~\cite{OpenLinkVirtuoso:webpage}) and integrates custom SPARQL
queries into the user interface e.g. to allow access to PlanetMath articles via the
MSC2010 classification~\cite{MSC2010}.

Already the proof-of-concept implementation in Fall 2010 made it clear that this
combination of MKM technologies could be much more useful than for just re-implementing
PlanetMath.

\subsection*{Framework for Semantic Publishing \& Active Documents}

\begin{wrapfigure}r{7.5cm}\vspace*{-2em}
\begin{tikzpicture}[scale=0.6,transform shape]\footnotesize
  \pgfdeclareimage[width=1cm]{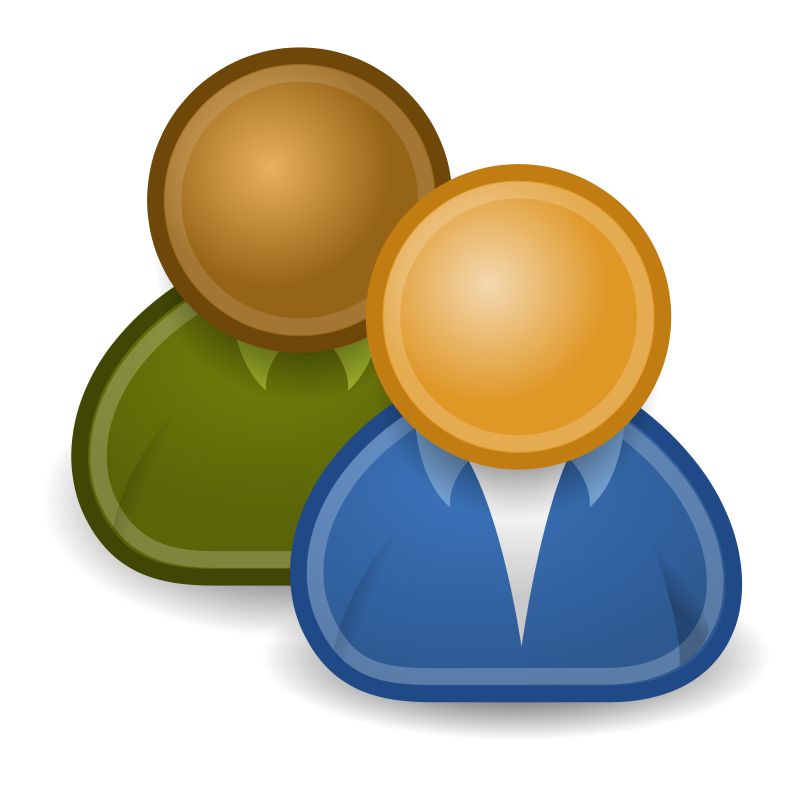}{user}
 
  \tikzstyle{system} = [rectangle, draw, fill=blue!20, text width=1.6cm, text centered,
                                    rounded corners, minimum height=1cm,shade, 
                                    top color=white, bottom color=blue!20]
  \tikzstyle{arrowleft} = [single arrow, draw, inner sep=1pt, anchor=west,
                                       minimum width=1cm, minimum height=1.7cm, 
                                       fill=lightgray!20,single arrow head indent=.4ex]
  \tikzstyle{arrowright} = [single arrow, shape border rotate=180, draw, inner sep=1pt,
                                         anchor=west,minimum width=1cm,minimum height=1.7cm,
                                         fill=lightgray!20,single arrow head indent=.5ex]
 \tikzstyle{arrowdown} = [single arrow,rotate=270, draw, inner sep=1pt,
                                         anchor=west,minimum width=1cm,minimum height=1.7cm,
                                         fill=lightgray!20,single arrow head indent=.5ex] 

\node[rectangle] (user) {\pgfuseimage{user}}; 
\node[system,right=1.7cm of user] (firefox) {Firefox}; 
\node[system,right=1.7cm of firefox] (drupal) {CMS};
\node[system,right=1.9cm of drupal] (tnt) {TNTBase};

\node[right=0.1cm of drupal, arrowright] (ar2) {HTML5};
\node[system, above=1.4cm of ar2] (latexml) {LaTeXML}; 
\node[system, below=1.4cm of ar2] (virtuoso) {Virtuoso};
\node[above right=0.1cm of firefox, arrowleft] (ar3) {REST}; 
\node[below right=0.1cm of firefox, arrowright] (ar4) {HTML5};

\node [double arrow, left=0.1cm of firefox,draw, 
            minimum height=1.7cm, minimum width=1cm, fill=lightgray!20, double arrow head indent=.5ex] {}; 
\node[below=0.1cm of drupal,double arrow,rotate=315, draw, inner sep=1pt, anchor=west,
           minimum width=1cm,minimum height=1.7cm,fill=lightgray!20,single arrow head indent=.5ex]
           {SPARQL}; 
\node[above=0.1cm of drupal,single arrow,rotate=45, draw, inner sep=1pt,anchor=west,
           minimum width=1cm,minimum height=1.7cm,fill=lightgray!20,single arrow head indent=.5ex] {\sTeX}; 
\node[below right=0.1cm of latexml,single arrow,rotate=315,draw,
          inner sep=1pt, anchor=west,minimum width=1cm,minimum
          height=1.7cm,fill=lightgray!20,single arrow head indent=.5ex] {OMDoc};
\node[below=0.1cm of tnt,single arrow,rotate=235, draw, inner sep=1pt,
           anchor=west,minimum width=1cm,minimum height=1.7cm,fill=lightgray!20,
           single arrow head indent=.5ex] {RDF}; 

\draw[->,thick] (firefox) to[loop below] node [below] (jobad) {JOBAD} (firefox); 
\draw[<->,dashed] (jobad) -- (virtuoso);
\draw[<->,dashed] (jobad) -- (tnt);
\node[rectangle,rounded corners,draw,inner sep=6pt,dotted, 
         fit=(drupal) (latexml) (virtuoso) (tnt)] (pacbox) {};
\node[above=1cm of tnt] {\hspace*{1em}\begin{tabular}{r}Content\\ Management\\ System\end{tabular}};
\end{tikzpicture}\vspace*{-1em}
\caption{Architecture of the \protect\pns}\label{fig:arch}\vspace*{-1em}
\end{wrapfigure}
The \pns has been generalized into a comprehensive eMath3.0 framework for semantic
publishing and knowledge management, which has been instantiated prototypically in a
variety of settings to validate the framework and support communities. The portals
realized in the \pns range from eLearning systems over scientific archives to semantic
help systems. All share the common basic architecture (see Figure~\ref{fig:arch}), which
integrates the components discussed above as web services into a central \emph{Container}
Management System (CMS) that mediates all user interaction. Note that in this MKM-centric
architecture, we greatly extend the role of the \emph{content management subsystem}
(denoted by the dotted box in Figure~\ref{fig:arch}). The CMS (initially Vanilla Forums,
later Drupal) supplies management and interaction at the ``container level'', i.e., without
ever looking into the documents it manages (hence the somewhat non-standard name). The
management of \emph{structured document content} is split between TNTBase and the RDF
triple store in \pn, since they can perform semantic services.

\subsection*{Active Documents by Example}

Note that the level of semantic interaction afforded by the \pns depends on the depth of
semantic annotations in the documents, and thus on different instances of the \pns: They
range from simple folding and localized commenting services in a front-end system for the
Cornell ePrint arXiv to a front-end system which features in-place type reconstruction and
elision of arguments and brackets for the fully formal LATIN
atlas~\cite{CodHorKoh:palai11}.

\begin{wrapfigure}r{.69\textwidth}\vspace*{-2em}
\includegraphics[width=.69\textwidth]{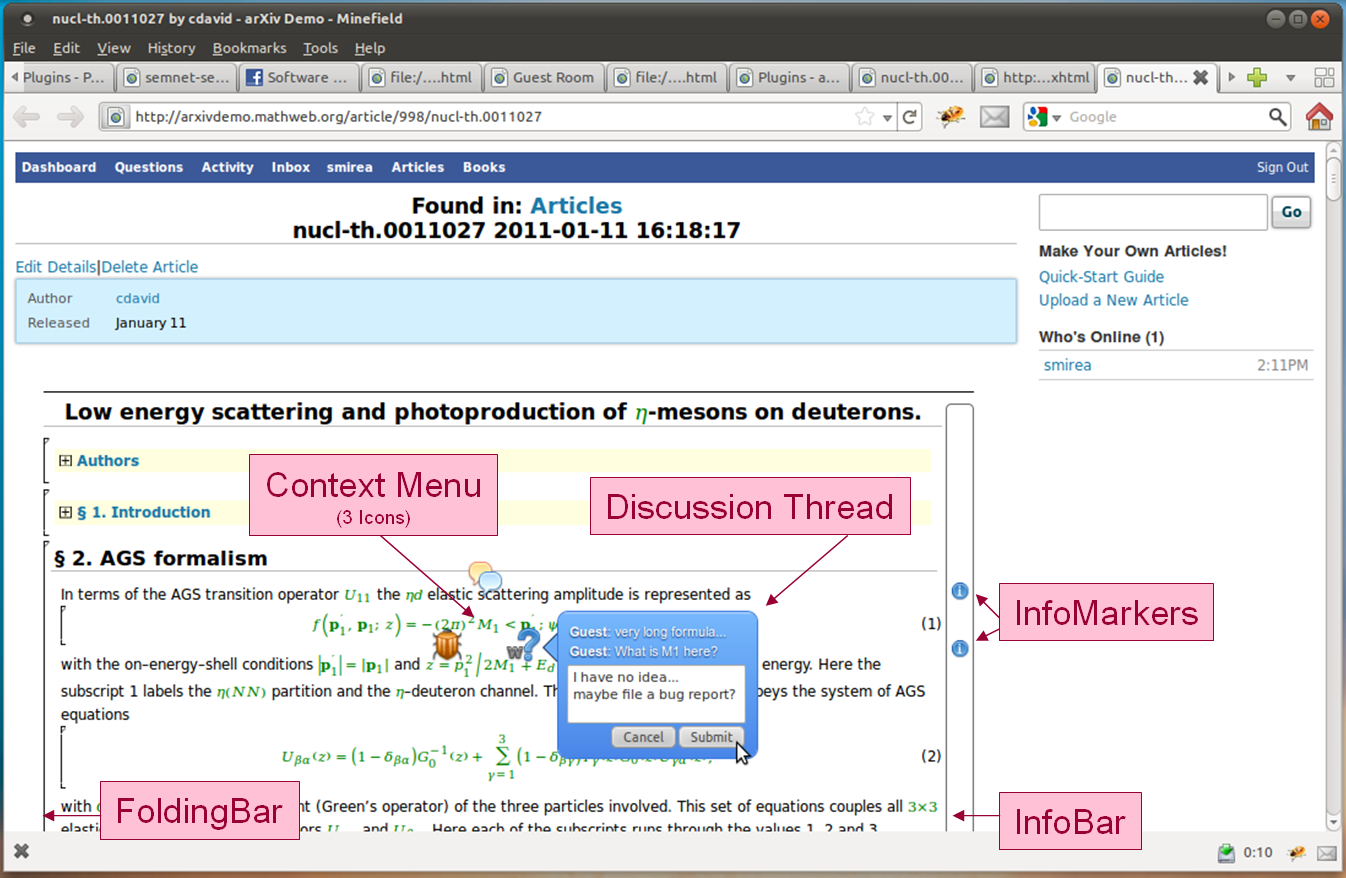}\vspace*{-.5em}
\caption{Interacting with an arXiv article}\label{fig:infobar}\vspace*{-1.5em}
\end{wrapfigure}
\looseness=-1\noindent We start with the former and work our way to more semantics. In all
cases, services are accessible locally for objects with (fine-grained) semantic
annotations -- e.g.  a subterm of a formula -- via a special context menu menu of icons
centered around the object.  The icon menu has one entry per service available in the
current context. For instance, the question mark icon triggers the discussion service
supports localized discussion threads fror reporting problems or asking questions about
the selected object. The “{\infobar}” on the right of Figure~\ref{fig:infobar} is a
secondary device that visualizes state information for the objects in the respective line
of the paper, e.g. the availability of questions or discussions, which can be accessed by
the icon menu.

\begin{wrapfigure}l{3cm}\vspace*{-2em}
  \begin{tabular}{c}
    \includegraphics[width=3cm]{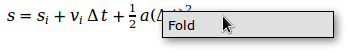}\\
    \includegraphics[width=3cm]{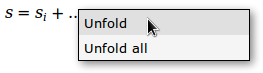}
  \end{tabular}\vspace*{-5ex}
\end{wrapfigure}
\noindent The \foldingbar in Figure~\ref{fig:infobar}, similar to source code IDEs,
enables folding document structures, and the \infobar icons on the right indicate the
availability of local discussions. In the image on the left, we selected a subterm and
requested to fold it, i.e. to simplify its display by replacing it with an ellipsis.

\begin{figure}\centering
  \begin{tabular}{|p{7cm}|p{5cm}|}\hline 
  \vspace{-5.5cm}\hspace{2cm}\fbox{\includegraphics[width=2cm]{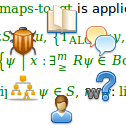}}\newline\vspace{-.25cm}\newline\includegraphics[width=5.5cm]{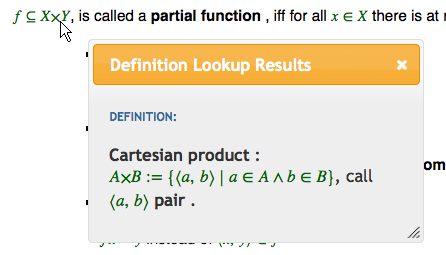} & 
  \includegraphics[width=5cm]{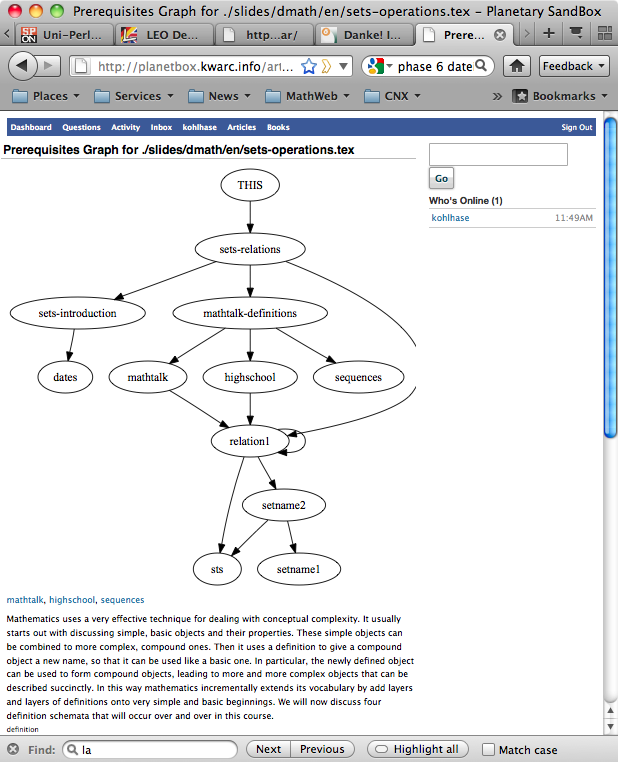}\\\hline
\end{tabular}
\caption{Definition Lookup and Prerequisites Navigation}\label{fig:definition-lookup}
\end{figure}
The richer semantic markup of \omdoc-based representations of lecture materials and the
Logic Atlas collections enable services that utilize logical and functional structures --
reflected by a different icon menu.  Figure~\ref{fig:definition-lookup} demonstrates
looking up a definition and exploring the prerequisites of a concept.  The definition
lookup service obtains the URI of a symbol from the annotation of a formula and queries
the server for the corresponding definition.  The server-side part of the prerequisite
navigation service obtains the transitive closure of all dependencies of a given item and
returns them as an annotated SVG graph.

\subsection*{Current  and Future Work}

In June 2011, the \pn system became one of the finalist systems in the Elsevier Executable
Papers Challenge~\cite{KohDavGin:psewads11}. But the development push to reach this
milestone also revealed crucial shortcomings of the CMS at the heart of the \pns, and the
system was ported to Drupal whose container model and editing facilities are more
suitable. Unfortunately, work on the port, on improving the subsystems, and data
conversion issues have delayed any deployment of production systems based on \pn

Currently, the work in the \pnp focuses on four \pn-based systems:
\begin{compactitem}
\item finishing a production-ready \pn instance of PlanetMath, see
  \url{http://alpha.planetmath.org}
\item developing a Web2.0 frontend with lightweight semantic features for
  \url{http://arxiv.org}, an archive of over $700\;000$ scientific documents.  Particular
  care will be placed on extracting functional semantics from give {\LaTeX} documents and
  using this in formula search, see \url{http://arxivdemo.mathweb.org}
\item re-establishing the separate compilation and linking functionality for modular
  semantic publishing (see~\cite{DavGin+:fmspscdl11}) in the eLearning3.0 System PantaRhei
  used in teaching CS courses at Jacobs University, see \url{http://panta.kwarc.info} and
\item integrating \pn as a knowledge provider in semantic allies; see
  \cite{DavJucKoh:safusa12}.
\end{compactitem}
Note that all the \pn instances referenced in the URIs are under active research, so your
experience may vary. 
\printbibliography
\end{document}